\documentstyle[12pt]{article}
\def\sepand{\rule{14cm}{0pt}\and}
\newcommand{\beq}{\begin{equation}}
\newcommand{\eeq}{\end{equation}}
\newcommand{\bq}{\begin{quotation}}
\newcommand{\eq}{\end{quotation}}
\newcommand{\bc}{\begin{center}}
\newcommand{\ec}{\end{center}}

\begin{document}

\title{QCD deconfinement -- phase transitions and collapsing  quark stars.}  
\vspace{0.6 cm}

\author{
{\sc Ilona Bednarek}\\
{\sl Institute of Physics,} \\
{\sl Silesian University,}\\
{\sl Uniwersytecka 4, 40-007 Katowice, Poland}  \\
\sepand
{\sc Marek Biesiada}\\
{\sl Copernicus Astronomical Center, } \\
{\sl Bartycka 18, 00-716 Warsaw, Poland}  \\
\sepand
{\sc Ryszard Ma\'{n}ka}\\
{\sl Institute of Physics,} \\
{\sl Silesian University,}\\
{\sl Uniwersytecka 4, 40-007 Katowice, Poland}  \\
}
\date{}
\maketitle
\vfill
\newpage
\begin{abstract}              
\noindent

In this paper we discuss 
the QCD phase-transitions in the nontopological soliton model of quark
confinement and explore possible astrophysical consequences.
Our key idea is to look at quark stars (which are believed to exist since the
quark matter is energetically preferred over the ordinary matter) from the
point of view of soliton model. 

We propose that the phase transition taking place during the core
collapse of massive evolved star may provide a new physical effect not taken
into account in modeling the supernova explosions. We also point out the
possibility that merging quark stars may produce gamma-ray bursts energetic
enough to be at cosmological distances. Our idea based on the
finite-temperature nontopologiocal soliton model overcomes major difficulties 
present in neutron star merger scenario --- the baryon loading problem and
nonthermal spectra of the bursts.

\end{abstract}
\vfill
\leftline{PACS number(s): 98.80.Cq, 12.15.Cc}
\newpage

\section{Introduction.}

The possibility that 
quark matter could be the ground state of hadronic
matter attracted a lot of attention since Witten's seminal paper \cite{Witten}.

There are generally two kinds of natural settings in which to seek for
macroscopic 
quark configurations. First of them stems from cosmological quark-hadron
phase-transition which occured when the temperature of the early Universe
dropped below $100$--$200$ MeV \cite{cosm}. 
The second one is of an astrophysical nature
and places quark matter inside compact objects like strange stars, cores of
neutron stars etc. \cite{strangestar}. 
These possibilities are by now rather well
studied (see \cite{Madsen} for a review). The basic 
physical model adopted in studies of the quark matter is the 
MIT bag model for hadrons. In this picture the quark star (or primordial quark
lump) is understood as an enormous "hadron" i.e. an ensamble of quarks
inside a confining bag. The notion of bag is believed to capture, at a
phenomenological level the essence of
(yet unknown precisely) zero-tempetrature nonperturbative QCD effects. 
The simplest way to generalize the MIT model to a finite temperature theory is
to adopt the Friedberg-Lee nontopological soliton model \cite{FLee}. In this
picture the "bag constant" is again a difference between the values of an
effective potential $U_{eff}$ in perturbative QCD vacuum and in physical vacuum
respectively. However the effecitve potential is temperature-dependent now and
so are its extrema. Hence the bag constant is no longer a constant but rather a
function of temperature $T$ (and chemical potential $\mu$). This opens a
possibility of phase-transitions in macroscopic quark configurations. 
Phase-transitions in a nontopological soliton model of
hadrons has been studied by Song,Enke and Jian Ong \cite{deconfinement} with
hope to understand the confinement. 
Our investigation performed in the next section is in a similar vein. We
demonstrate that macroscopic quark configurations may undergo phase transitions
at some critical temperature releasing thereby a huge amount of energy. 
Then in section 3 we contemplate 
possible astrophysical consequences of such phase-transitions. 
Section 4 contains concluding remarks.

\section{Phase transitions in Friedberg-Lee quark solitons.}

We focus our attention on the Friedberg-Lee nontopological soliton model 
defined by the Lagrangian density
\cite{FLee}
\begin{eqnarray} \label{lagrangian}
{\cal L} = 
\frac{1}{2} \partial_{\mu} \varphi\partial^{\mu} \varphi-U(\varphi)  
 + \overline{\psi} (i\gamma^{\mu}D_{\mu} - m - g_0 \varphi) \psi \nonumber 
\end{eqnarray}
where $\psi$ is the quark - field, $m$ -- the quark mass and $\varphi$ is a
phenomenological scalar field believed to represent nonperturbative QCD
effects. 
The self-interaction  
potential of the scalar field $\varphi$ is assumed to be of the form: 
\begin{equation} \label{potential}
U(\varphi) = \frac{a}{2!}\varphi^{2} - \frac{b}{3!}\varphi^{3} +
\frac{c}{4!}\varphi^{4} + B.
\end{equation}
The constants $a, b, c$ and $g_{0}$ represent four parameters of the model. 
The potential $U(\varphi)$ has typically two minima, one at $\varphi_0 = 0$
which is a local minimum associated with the perturbative vacuum
state and the second --- absolute minimum, at
\begin{equation} \label{physvac} 
\varphi_{\upsilon} = \frac{3|b|}{2c} \left[ 1 + (1 -
\frac{8ac}{b^2})^{\frac{1}{2}} \right]
\end{equation}
corresponding to the physical vacuum.
The constant $B$ measures the difference in potential between two vacuum states
and thus has the same meaning as the MIT bag constant. 
However unlike in the MIT model the confining bag is not put in {\sl \'a
priori} but 
rather results from the dynamics of the scalar field $\varphi.$
 
Due to nonlinearity of the potential $U(\varphi)$ the Lagrangian
(\ref{lagrangian}) 
leads to nontopological soliton solutions. 
There are two limiting types of the shape of $U(\varphi)$ for which the soliton
solution exists.
The first represents degenerate vacuum state with both minima at the same level
$U(\varphi_{\upsilon}) = U(\varphi_0)$ 
and consequently $B = 0$. 
The second case is the one 
when the local minimum and maximum overlap producing thus 
an inflection point.  This is the moment when the perturbative
and physical vacuum states become metastable. Soliton solutions exist for all
intermediate potential shapes i.e. for $U(\varphi)$ with two minima and one
maximum. At this stage the demand that soliton solution exist imposes a
restriction on the $a, b, c$ -- parameters of the model. 

It is now interesting to study how the properties of soliton solutions change 
with temperature. 
For this purpose it is convenient to perform the decomposition of the scalar
field $\varphi$ into thermodynamical average 
$$
\sigma = <\varphi> = \frac{Tr e^{-\beta H}\varphi}{Tr e^{-\beta H}}
$$
and quantum corrections $\tilde \varphi$:
\begin{equation} \label{decomposition} 
\varphi = \sigma + \tilde \varphi.
\end{equation}
Consequently the potential function $U(\varphi)$ transforms into an effective
potential $U^{\beta}_{eff}(\sigma)$ which is explicitly temperature-dependent: 
\begin{equation} \label{effective} 
U^{\beta}_{eff}(\sigma) = U^{T = 0}_{eff}(\sigma) +
\delta_{\beta}U_{eff}(\sigma) 
\end{equation}
where $U^{T = 0}_{eff}(\sigma)$ is the zero-temperature one-loop effective
potential and the one-loop temperature dependent quantum correction reads: 
\begin{eqnarray} \label{one-loop} 
\delta_{\beta}U_{eff}(\sigma) & = & \frac{1}{\beta}\int
\frac{d^{3}k}{(2\pi)^{3}}ln(1-e^{-\beta E_{M}})  \nonumber \\ 
                            & = &
\frac{1}{2\pi^{2}\beta^{4}} \int_{0}^{\infty} x^{2} dx ln(1-e^{-(x^{2} +
\beta^{2}M^{2})^{\frac{1}{2}}}) 
\nonumber \\ 
\end{eqnarray}
where $x =\beta k $, $k$ is the wave number, $E_{M}^{2}= k^{2} + M^{2}$ and  
$M$ is the effective mass of the scalar field i.e. equal to the second
derivative of $U^{\beta}_{eff}(\sigma)$ taken at the global minimum. 

The expression (\ref{one-loop}) is in a sense incomplete since the
contributions coming from fermions have been neglected. However such thermal
corrections are suppossed to be significant only at temperatures exceeding the
Fermi level $kT >> E_F$ and an inspection at Table 2 shows that 
indeed 
Fermi energies are of order of several hundreds of MeV
which is much greater than thermal energies encountered during the core
collapse. 

Expanding $\delta_{\beta}U_{eff}(\sigma)$ in 
power series around $\beta M =0$ in the integral (\ref{one-loop})
yields the well known expresion 
\begin{eqnarray}
\delta_{\beta}U_{eff}(\sigma) & = &
\frac{1}{2\pi^{2}\beta^{4}} ( \int_{0}^{\infty} x^{2}dx ln(1-e^{-x})
\nonumber \\
& + & \beta^{2}M^{2}\frac{1}{2}\int _{0}^{\infty}
dx\frac{x}{(e^{x}-1)} + O (\beta^{4}M^{4}) ) \nonumber \\ 
\end{eqnarray}
which can be integrated to give the following formula: 
\begin{equation}
\delta_{\beta}U_{eff}(\sigma) = -\frac{\pi ^{2}}{90}T^{4} +
\frac{T^{2}}{24}M^{2}. 
\end{equation}
Therefore the effective potential $U_{eff}(\sigma,T)$ is equal to
\begin{eqnarray} \label{Ueffective} 
U_{eff}(\sigma,T) & = & -\frac{bT^{2}}{24}\sigma + \frac{1}{2}(a +
\frac{cT^{2}}{24})\sigma^{2} - \frac{1}{6}b\sigma^{3}   \nonumber \\ 
                & + & \frac{c}{24}\sigma^{4} + \frac{a}{24}T^{2} +
\frac{c}{1152}T^{4} - \frac{\pi^{2}}{90}T^{4} + B  \nonumber \\ 
\end{eqnarray}
The temperature dependence of the effective potential $U_{eff}(\sigma,T)$ is
illustrated on Figure 1.
For $T = 0$  the effective potential has three extrema
corresponding to the perturbative vacuum, local maximum (potential barrier) and
physical vacuum, respectively.
At some critical temperature $T_{0}$ located here somewhere between 
$0.3\;m_s$ and $0.4\;m_s$, where $m_s = 197.32$ MeV is the strange quark mass, 
the perturbative vacuum and the local maximum overlap, inflection point
develops and bag constant vanishes.  This is also the last moment when the
soliton exists. 
Values of the temperature $T_{0}$ for each set of parameters are presented in
Table 1. \\

\begin{table}[h]
\centering
\begin{tabular}{|l|r|r|r|l|}
\hline
        & a($fm^{-2}$) & b($fm^{-1}$)  & c       &  $T_{0}$(MeV)  \\ \hline
{\bf 1} & 51.60        & 799.90        & 4000    &  68.71          \\ \hline 
{\bf 2} & 7.671        & 107.27        & 500     &  84.41         \\ \hline 
{\bf 3} & 12.85        & 196.34        & 1000    &  77.28          \\ \hline 
{\bf 4} & 40.88        & 783.80        & 5000    &  60.46           \\ \hline 
{\bf 5} & 66.42        & 1411.60       & 10000   &  55.59         \\ \hline 
{\bf 6} & 107.32       & 2537.00       & 20000   &  49.52          \\ \hline 
{\bf 7} & 321.75       & 9824.80       & 100000  &  38.67          \\ \hline 
\end{tabular}
\caption{Values of the critical temperature $T_{0}$ for selected set of
parameters.}  
\end{table}

\begin{table}
\begin{center}
\label{valueF}
\vspace{0.5cm}
\begin{tabular}{|c||c||c|}
\hline
 & B & $k_F$\\ \hline \hline
ad {\bf 1}& $(100 MeV)^4$ & 277 MeV \\ \cline{2-3}
& $(180 MeV)^4$ & 499.3 MeV \\ \hline \hline
ad {\bf 2}& $(100 MeV)^4$ & 306.9 MeV \\ \cline{2-3}
& $(180 MeV)^4 $ & 552.6 MeV \\ \hline
\end{tabular}
\caption{Fermi levels for selected model parameters.}
\end{center}
\end{table}

Alternative way of illustrating this transition is presented in Figure 2 where
the location of extrema of the effective potential is plotted against the
temperature. 

For temperatures greater than $T_{0}$
the soliton disappears and the structure of the vacuum changes.
By comparing the effective potential (\ref{Ueffective}) with the potential
(\ref{potential}) one may identify last four terms in (\ref{Ueffective}) with
temperature dependent "bag constant"  $B(T)$ describing
the pressure of the physical vacuum exerted on the false vacuum 
\begin{equation}
B(T) = \frac{a}{24}T^{2} +
\frac{c}{1152}T^{4} - \frac{\pi^{2}}{90}T^{4} + B . \nonumber \\ 
\end{equation}
Fig.3 shows $ B(T)$ as a function of temperature calculated for the set of
parameters no.1 from Table 1. At the critical temperature $B(T)$ is nonzero 
(equal to $\sim 0.1 \times m_s^4$ in this specific example) despite the fact
that the soliton disappears. It means that the energy contained in the bag
soliton is liberated at the critical temperature. This observation is crucial
for possible astrophysical scenario envisaged in this paper. 

Before closing this section let us discuss the behavior of the total energy of
the quark core. 
The total energy of a soliton bag (including  self-gravity) 
takes the following form 
\begin{equation} \label{total}
E =\frac{\alpha N^{\frac{4}{3}}}{R} +\frac{4}{3}\pi BR^{3}
-\frac{3}{5}\frac{G}{R}(\frac{\alpha N^\frac{4}{3}}{R} +\frac{4}{3}\pi
BR^{3})^{2}, 
\end{equation}
where $\alpha = \frac{9}{4}(\frac{9 \pi}{4})^{1/3}$.
Following \cite{Cottingham} it is convenient to introduce a dimensionless 
quantity
\begin{equation} \label{x}
x=(\frac{4\pi B}{\alpha})^{\frac{1}{4}}N^{\frac{-1}{3}}R
\end{equation}
Then one has, 
\begin{equation} \label{Ex} 
E = \alpha^{\frac{3}{4}}(4B\pi)^{\frac{1}{4}}N(\frac{1}{x} + x^{3} -
\frac{e}{x} (\frac{3}{x}+x^{3})^{2}) 
\end{equation}
with
\begin{equation}
e = \frac{1}{5}GN^{\frac{2}{3}}(4\pi \alpha B)^{\frac{1}{2}}.
\end{equation}
Figure 4 illustrates the dimensionless energy function $f(x)$
\begin{equation}
f(x) = \frac{1}{x}+x^{3}-\frac{e}{x}(\frac{3}{x}+x^{3})^{2}
\end{equation}
represented by a solid curve. 
It has a local minimum corresponding to the stable core configuration where the
repulsive term coming from the Fermi statistics obeyed by the quarks is
counterbalanced by the joint action of bag pressure and gravity. 
We stress that such equilibrium configuration may not be realised during the
collapse which is a rapidly progressing dynamical process. One may imagine that
while the core is collapsing one is moving along the curve $f(x)$ to the left. 
When the effective bag constant
$B(T)$ vanishes the bag pressure vanishes as well and only much weaker
effect of gravity remains to act against fermion repulsion.
This changes qualitatively the $f(x)$-function and
the new situation is denoted by curve $f_{0}(x)$ in Figure 3.
Hence at the moment when $B(T)$ vanishes
the point representing the energetic state of the core jumps (at some $x_{cr}$)
from the curve $f(x)$ to $f_0(x)$ liberating some energy in this process. 
Moreover the repulsive, fermionic term is dominant now and causes the expansion
of the core.  

\section{Pos\-sible as\-tro\-phy\-si\-cal con\-se\-quen\-ces of
high-tem\-pera\-ture QCD phase--transitions.} 

The history of science in the present century (or at least its second half) 
reveals 
a successful search for the manifestation of
fundamental interactions in astrophysical and cosmological context.
Consequently for any branch of physics one may find a corresponding branch of
astrophysics \cite{Hu}. With this perspective in mind we outline in this
section some astrophysical processes in which the QCD effects discussed above
are likely to be operating.

\subsection{Core collapse and the supernova explosion.} 

Massive stars are known to end their evolution with the destabilization and
collapse of their iron cores. Following the core collapse a supernova explosion
releases most of the matter (and the internal energy) of the progenitor. 
Although the phenomenon of supernova is known (observationally) since ancient
times and despite the tremendous progress achieved in the theory of stellar
evolution (especially in the field of numerical simulations) the detailed
mechanism of explosion is not well understood. 

The most commonly accepted picture is the following: the destabilization of the
iron core leads to core collapse during which the core divides into a subsonic,
homologously collapsing inner part and outer free-falling part. The collapse of
the inner part is halted at nuclear densities by the stiffening of the equation
of state and after a rebounce an outwardly propagating shock wave is generated.
Numerical simulations show that usually the shock wave is damped mainly due to
photodisintegration of Fe-like elements and becomes a standing accretion shock.
As a way out of this trouble it has been proposed by Wilson \cite{Wilson} that
stagnated 
shock may be revived by the accretion induced heating due to neutrino
absorbtion. 
Simple criterion for the shock revival can be established in terms of neutrino
luminosity $L_{\nu}$ and the average of the rms $\nu_e$ energies
$\langle \varepsilon_{\nu}^2 \rangle$ \cite{Bruenn}:

\begin{equation} \label{revival} 
L_{51} \langle \varepsilon_{15}^2 \rangle >  48
\frac{M_{1.4}^{3/2}}{r_3^{1/2}},  
\end{equation}
\noindent
where:$L_{51} =\displaystyle{ \frac{L_{\nu}}{10^{51}\;erg\;s^{-1}}},$ 
$\displaystyle{\varepsilon_{15} =\frac{\varepsilon_{\nu}}{15\;MeV}},$ 
$M_{1.4}=\displaystyle{\frac{M}{1.4\;M_{\odot}}}$ and $r_3 =
\displaystyle{\frac{r}{10^3\;cm}}.$ 
It means that $L_{51} \langle
\varepsilon_{15}^2 \rangle$ should be large enough in order to revive the
shock. This is not 
always easy to acheive in a way proposed by Wilson i.e. by neutrino absorbtion.
However any other mechanism
for the revival is good provided the heating rate exceeds the value 
\begin{equation} \label{heatingrate}
\frac{d \varepsilon}{dt} = 53 \frac{M_{1.4}^{3/2}}{r_3^{1/2}}\;\;\;MeV\;s^{-1}.
\end{equation}
The need for a robust 
mechanism of explosion is even more pronounced in the context
of the Supernova 1987A because the progenitor here was an 18--20 $M_{\odot}$ 
blue supergiant and it is commonly accepted that the prompt shock wouldn't work
in such massive star \cite{Brown}. 

Various phase transitions are expected to occur in the core. Above the nuclear
density there may be first or second order phase transitions such like pion and
kaon condenstes \cite{Brown} or deconfined quark matter \cite{Benvenuto}.
There were attempts to include this complicated physics into numerical
simulations of the core collapse but the results are clearly still
controversial in the aspect of explosion. Although some progress has recently 
been achieved in hydrodynamical treatment of neutrino transport 
it cannot be
excluded that physical effects outlined in this paper may find application in
realistic models of core collapse.

A proto-neutron star core is a favorable environment for the conversion of
ordinary hadronic to strange quark matter through a variety of fluctuations 
\cite{strangestar}. 
It seems reasonable to contemplate the properties of the quark matter in the
collapsing core within the finite-temperature formalism of Friedberg and Lee. 
The results of Cooperstein \cite{Cooperstein} are encouraging in this respect
since his simulations 
demonstrated that the temperature of collapsing core may easily acquire
$80\;MeV$ or higher. One may thus expect the occurence of phase transitions
discussed in previous section as an additional trigger for supernova explosion.
As shown in Fig.3 the energy density that may be liberated at the critical
point is of order of $\sim 0.09\;(m_s)^4 \approx 197.33\;MeV/fm^3$ and assuming
the radius of the quark core of order of $\sim 10\;km$ this gives the total
energy equal to $\sim 10^{53}\;erg$.

\subsection{Gamma-bursts from strange stars' mergers.}

Since their discovery about 20 years ago \cite{KSO} 
gamma ray bursts (GRBs) still remain
the most mysterious phenomena in the sky \cite{Paczynski}. 
Recent results provided by the
ongoing mission of the Compton GRO \cite{GRO} demonstrated an isotropic
distribution of GRBs on the celestial shpere which combined with their radial
distribution being uniform out to a certain distance and falling off beyond 
strongly supported the idea that GRBs
originate from cosmological sources \cite{cosmological}. 
More recently the correlation between the duration
the strength and the hardness of the bursts has been found \cite{tdist}
as predicted by the cosmological scenario. 
In the light of enormous diversity of durations, time variability and
spectra of the GRBs no detailed theoretical model gained general
acceptance yet. 
There is however a number of reasons for believing that the GRBs may come from
neutron star mergers. Neutron star binaries such like the famous binary pulsar
PSR 1916+13 are known to exist and because they radiate away gravitational
waves as predicted by general relativity they will finish their lifes in a
catastrophic 
merger event powerful enough to be responsible for GRBs at cosmological
distances.  
The expected rate of neutron star mergers (inferred from the known three binary
pulsars) is about $10^{-5.5 \pm .5}\;{\rm /year/galaxy}$ \cite{Tsvi}. 
This is in perfect agreement with a local rate of $\sim 10^{-6}\;{\rm
/year/galaxy}$ seen in the BATSE experiment. 

The idea that the strange quark matter is the ground state of matter suggests 
that neutron stars may in fact be hybrid stars consisting of quark central
region covered by a layer of ordinary matter \cite{strangestar,Madsen}. 
Consequently the binary systems made of quark stars may exist as well. 
The idea that the GRBs could be explained by the collision of
two strange stars has been put foreward by Haensel, Paczy{\'n}ski and
Amsterdamski \cite{HPA}. Their line of thinking was following: a merger of two
strange stars thermalizes the kinetic energy of the collision which is roughly
about $10^{53}\;{\rm ergs.}$ then the newly born post-merger starnge star
radiates most of the thermal energy in the form of neutrinos and antineutrinos
within a few seconds after a collision. The neutrino-antineutrino pairs
convert into electron-positron pairs which annihilate into gamma rays. Because
the huge amount of energy is deposited in a small region (surface of
post-merger star) on a time scale of a fraction of a second the $e^+,e^-$
plasma  will be thermalized and will produce a fireball with a blackbody
spectrum peaking at energies $\sim 10 \;{\rm MeV}.$ The fireball model faces
two serious problems: the effect of baryons and the origin of the observed
nonthermal spectrum of GRBs. The baryon loading 
resulting in optically thick fireball and subsequent conversion of radiation
into kinetic energy of the fireball 
is naturally overcome in the
case of quark stars (no baryons) but the second objection remains. 

Let us consider the merger of two quark stars from the point of view of
nontopological soliton approach presented in the section 2. 
Assume that they are initially cool stable configurations which masses and
radii are determined by the bag constant $B:$ \cite{Cottingham}
\begin{eqnarray} \label{M,R}
M_{1,2}(B) &=& \mu B^{1/4} N_{1,2}  \nonumber \\
R_{1,2}(B) &=& \rho B^{-1/4} N_{1,2}^{1/3} \nonumber
\end{eqnarray}
where $\mu = 4 (\frac{3}{2})^{2/3} \pi^{1/2},$ $N$ is the total number of
fermions (quarks) and $\rho = \frac{1}{8}
(\frac{3}{2})^{5/3} \pi^{1/12}.$

For the post-merger object one should not take zero-temperature bag
constant $B$ but rather 
the effective bag constant  $B(T_{pm})$ where
the post-merger temperature $T_{pm}$ is of order of 
$10^{11}\;K$  \cite{HPA} 
since the kinetic energy of the collision is thermalized. Therefore the mass of
the final configuration $M_{pm} = \mu B(T_{pm})^{1/4} (N_1+N_2)$ is lower than
the sum of initial masses $M_1 +M_2$ and this mass deficit 
$\Delta M_{pm} = \mu
[B(T_{pm})^{1/4} -B(T=0)^{1/4}](N_1+N_2)$ 
is radiated away resembling somewhat 
the process of nuclear fussion. The phenomenological scalar field $\sigma$
undergoes out of equilibrium decay into radiation via a one-loop-fermion
process $\sigma \rightarrow 2\gamma$  \cite{Ng}.
Moreover it is oversimplified to assume that the post-merger configuration is
stable against collapse. 
This opens the possibility that the phase-transitions described in the Section
2 may be operating in this scenario as well. 
It should be stressed that the second-order phase transition which is seen in
the behavior of the effective bag constant $B(T)$ is closely related to the
behavior of the phenomenological scalar field $\sigma.$ 
At the critical point the scalar 
field $\sigma$ (or rather its excess over the hadronic matter case)
should be radiated away in a process $\sigma \rightarrow 2\gamma.$ 
This radiation would be nonthermal and could explain the nonthermal spectra of
the GRBs which are otherwise hard to explain. The detailed calculations of 
spectra in the merging quark stars scenario will be presented in another paper.

\section{Conclusions}

If one takes the idea of quark matter seriously then it is natural to 
consider the collapsing core of a massive evolved star as a prime site for the
quark matter formation. Because the temperatures typical for this environment
are high one concludes that the finite-temperature theory is more appropriate
than zero-temperature MIT model widely used in the literature. 
Consequently one may expect that massive compact objects produced in the final
stages of stellar evolution may contain quark matter (the so called hybrid
stars) or be composed exclusively out of it. If one adopts the philosophy that 
formation of degenerate gas provides the main force stabilizing the remnants of
stellar 
evolution against self-gravity, then we acquire a new link in the chain of
stable configurations:
 white dwarfs (degenerate electrons) $\rightarrow$ neutron stars
(degenerate neuron gas) $\rightarrow$ quark stars (degenerate quark gas) 
$\rightarrow$ black hole .

In this paper we tried to look at the problem of macroscopic quark
configurations from the point of view of finite-temperature theory --- the so
called Friedberg-Lee soliton model. We have demonstrated the possibility of a
phase transition during which the confining bag disappears thus liberating a
huge amount of energy. This may occur at temperatures of order of $\sim 40-80\;
MeV \approx 5-9\times 10^{11}\;K$ which is a range accessible during the core
collapse. The aforementioned phase transition may provide an important physical
effect 
(additional source of energy?)
during the 
supernova explosion. 
We have also presented an idea according to which merging quark stars discussed
in the finite-temperature theory may explain the gamma-ray bursts. Such a model
would easily overcome main difficulties that plagued conventional models such
like baryon loading and production of nonthermal radiation out of a thermalized
fireball.

\section*{Acknowledgements}

This project was sponsored by the KBN Grant 2 P304 022 06.

\newpage

\begin{center}
{\bf Figure captions}:
\end{center}
Figure 1. \\
Temperature dependence of the effective potential $U_{eff}(x,t),$ 
$(t = \displaystyle{\frac{T}{s_{0}}}$ and $x=\displaystyle{\frac{\sigma}{s_0}}$
where $s_{0} = 197.32$ MeV is the strange quark mass.)\\ 

Figure 2. \\
Values of $\sigma$ at the extremum of the effective potential plotted vs.
temperature -- the 
qualitative change from three to one extremum occurs at critical temperature.\\

Figure 3. \\
The temperature dependence of the bag constant $B(t)$\\

Figure 4. \\
The effective energy of the bag loaded with quarks $f(x)$,
$f_b(x)$ - the energy of the empty bag, $f_o(x)$ - the energy of free quarks.\\

\newpage
\begin{figure}
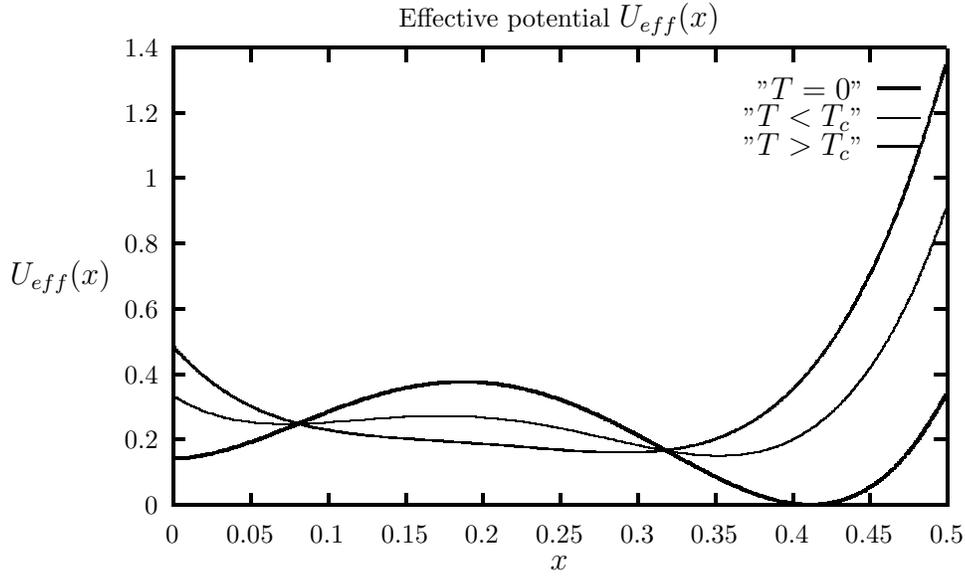

\input pot.tex
\caption{Temperature dependence of the effective potential $U_{eff}(x,t),$
$(t = \displaystyle{\frac{T}{s_{0}}}$ and $x=\displaystyle{\frac{\sigma}{s_0}}$
where $s_{0} = 197.32$ MeV is the strange quark mass.)}
\end{figure}
\newpage
\begin{figure}
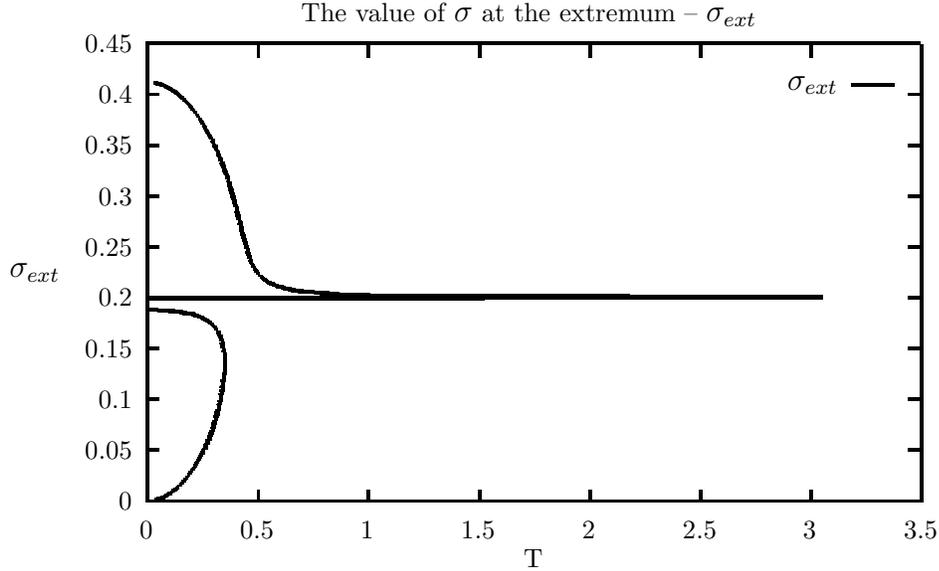

\input ord.tex
\caption{Values of $\sigma$ at the extremum of the effective potential plotted
vs. temperature -- the 
qualitative change from three to one extremum occurs at critical temperature.}
\end{figure}
\newpage
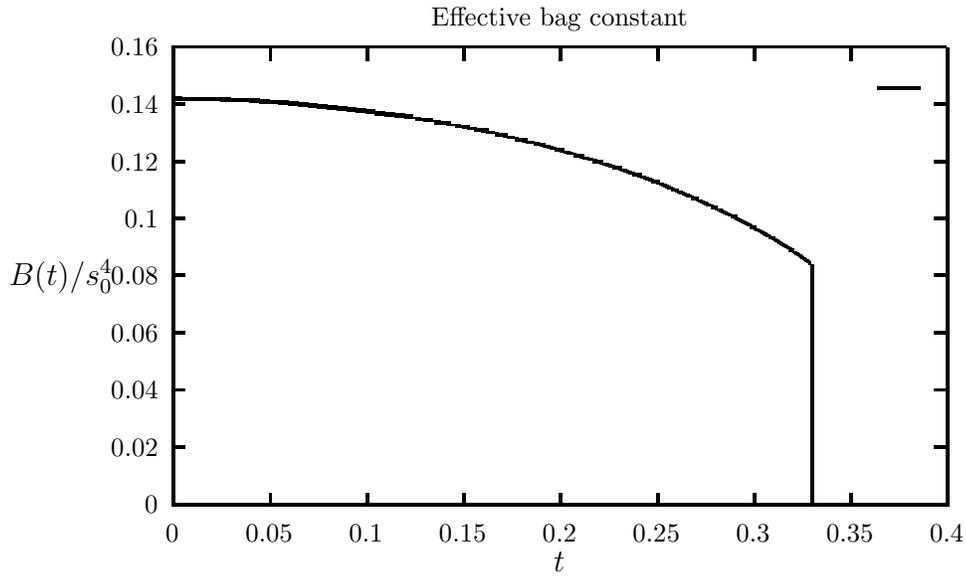
\begin{figure}
\setlength{\unitlength}{0.240900pt}
\ifx\plotpoint\undefined\newsavebox{\plotpoint}\fi
\sbox{\plotpoint}{\rule[-0.500pt]{1.000pt}{1.000pt}}%
\begin{picture}(1500,900)(0,0)
\font\gnuplot=cmr10 at 10pt
\gnuplot
\sbox{\plotpoint}{\rule[-0.500pt]{1.000pt}{1.000pt}}%
\put(220.0,113.0){\rule[-0.500pt]{292.934pt}{1.000pt}}
\put(220.0,113.0){\rule[-0.500pt]{1.000pt}{173.207pt}}
\put(220.0,113.0){\rule[-0.500pt]{4.818pt}{1.000pt}}
\put(198,113){\makebox(0,0)[r]{0}}
\put(1416.0,113.0){\rule[-0.500pt]{4.818pt}{1.000pt}}
\put(220.0,203.0){\rule[-0.500pt]{4.818pt}{1.000pt}}
\put(198,203){\makebox(0,0)[r]{0.02}}
\put(1416.0,203.0){\rule[-0.500pt]{4.818pt}{1.000pt}}
\put(220.0,293.0){\rule[-0.500pt]{4.818pt}{1.000pt}}
\put(198,293){\makebox(0,0)[r]{0.04}}
\put(1416.0,293.0){\rule[-0.500pt]{4.818pt}{1.000pt}}
\put(220.0,383.0){\rule[-0.500pt]{4.818pt}{1.000pt}}
\put(198,383){\makebox(0,0)[r]{0.06}}
\put(1416.0,383.0){\rule[-0.500pt]{4.818pt}{1.000pt}}
\put(220.0,473.0){\rule[-0.500pt]{4.818pt}{1.000pt}}
\put(198,473){\makebox(0,0)[r]{0.08}}
\put(1416.0,473.0){\rule[-0.500pt]{4.818pt}{1.000pt}}
\put(220.0,562.0){\rule[-0.500pt]{4.818pt}{1.000pt}}
\put(198,562){\makebox(0,0)[r]{0.1}}
\put(1416.0,562.0){\rule[-0.500pt]{4.818pt}{1.000pt}}
\put(220.0,652.0){\rule[-0.500pt]{4.818pt}{1.000pt}}
\put(198,652){\makebox(0,0)[r]{0.12}}
\put(1416.0,652.0){\rule[-0.500pt]{4.818pt}{1.000pt}}
\put(220.0,742.0){\rule[-0.500pt]{4.818pt}{1.000pt}}
\put(198,742){\makebox(0,0)[r]{0.14}}
\put(1416.0,742.0){\rule[-0.500pt]{4.818pt}{1.000pt}}
\put(220.0,832.0){\rule[-0.500pt]{4.818pt}{1.000pt}}
\put(198,832){\makebox(0,0)[r]{0.16}}
\put(1416.0,832.0){\rule[-0.500pt]{4.818pt}{1.000pt}}
\put(220.0,113.0){\rule[-0.500pt]{1.000pt}{4.818pt}}
\put(220,68){\makebox(0,0){0}}
\put(220.0,812.0){\rule[-0.500pt]{1.000pt}{4.818pt}}
\put(372.0,113.0){\rule[-0.500pt]{1.000pt}{4.818pt}}
\put(372,68){\makebox(0,0){0.05}}
\put(372.0,812.0){\rule[-0.500pt]{1.000pt}{4.818pt}}
\put(524.0,113.0){\rule[-0.500pt]{1.000pt}{4.818pt}}
\put(524,68){\makebox(0,0){0.1}}
\put(524.0,812.0){\rule[-0.500pt]{1.000pt}{4.818pt}}
\put(676.0,113.0){\rule[-0.500pt]{1.000pt}{4.818pt}}
\put(676,68){\makebox(0,0){0.15}}
\put(676.0,812.0){\rule[-0.500pt]{1.000pt}{4.818pt}}
\put(828.0,113.0){\rule[-0.500pt]{1.000pt}{4.818pt}}
\put(828,68){\makebox(0,0){0.2}}
\put(828.0,812.0){\rule[-0.500pt]{1.000pt}{4.818pt}}
\put(980.0,113.0){\rule[-0.500pt]{1.000pt}{4.818pt}}
\put(980,68){\makebox(0,0){0.25}}
\put(980.0,812.0){\rule[-0.500pt]{1.000pt}{4.818pt}}
\put(1132.0,113.0){\rule[-0.500pt]{1.000pt}{4.818pt}}
\put(1132,68){\makebox(0,0){0.3}}
\put(1132.0,812.0){\rule[-0.500pt]{1.000pt}{4.818pt}}
\put(1284.0,113.0){\rule[-0.500pt]{1.000pt}{4.818pt}}
\put(1284,68){\makebox(0,0){0.35}}
\put(1284.0,812.0){\rule[-0.500pt]{1.000pt}{4.818pt}}
\put(1436.0,113.0){\rule[-0.500pt]{1.000pt}{4.818pt}}
\put(1436,68){\makebox(0,0){0.4}}
\put(1436.0,812.0){\rule[-0.500pt]{1.000pt}{4.818pt}}
\put(220.0,113.0){\rule[-0.500pt]{292.934pt}{1.000pt}}
\put(1436.0,113.0){\rule[-0.500pt]{1.000pt}{173.207pt}}
\put(220.0,832.0){\rule[-0.500pt]{292.934pt}{1.000pt}}
\put(45,472){\makebox(0,0){$B(t)/s_0^4$}}
\put(828,23){\makebox(0,0){$t$}}
\put(828,877){\makebox(0,0){Effective bag constant}}
\put(220.0,113.0){\rule[-0.500pt]{1.000pt}{173.207pt}}
\put(1328.0,767.0){\rule[-0.500pt]{15.899pt}{1.000pt}}
\put(220,751){\usebox{\plotpoint}}
\put(220,748.42){\rule{7.227pt}{1.000pt}}
\multiput(220.00,748.92)(15.000,-1.000){2}{\rule{3.613pt}{1.000pt}}
\put(281,747.42){\rule{7.227pt}{1.000pt}}
\multiput(281.00,747.92)(15.000,-1.000){2}{\rule{3.613pt}{1.000pt}}
\put(311,746.42){\rule{7.468pt}{1.000pt}}
\multiput(311.00,746.92)(15.500,-1.000){2}{\rule{3.734pt}{1.000pt}}
\put(342,744.92){\rule{7.227pt}{1.000pt}}
\multiput(342.00,745.92)(15.000,-2.000){2}{\rule{3.613pt}{1.000pt}}
\put(372,742.92){\rule{7.227pt}{1.000pt}}
\multiput(372.00,743.92)(15.000,-2.000){2}{\rule{3.613pt}{1.000pt}}
\put(402,740.42){\rule{7.468pt}{1.000pt}}
\multiput(402.00,741.92)(15.500,-3.000){2}{\rule{3.734pt}{1.000pt}}
\put(433,737.42){\rule{7.227pt}{1.000pt}}
\multiput(433.00,738.92)(15.000,-3.000){2}{\rule{3.613pt}{1.000pt}}
\put(463,734.42){\rule{7.468pt}{1.000pt}}
\multiput(463.00,735.92)(15.500,-3.000){2}{\rule{3.734pt}{1.000pt}}
\put(494,730.92){\rule{7.227pt}{1.000pt}}
\multiput(494.00,732.92)(15.000,-4.000){2}{\rule{3.613pt}{1.000pt}}
\put(524,726.92){\rule{7.227pt}{1.000pt}}
\multiput(524.00,728.92)(15.000,-4.000){2}{\rule{3.613pt}{1.000pt}}
\put(554,722.92){\rule{7.468pt}{1.000pt}}
\multiput(554.00,724.92)(15.500,-4.000){2}{\rule{3.734pt}{1.000pt}}
\multiput(585.00,720.71)(4.056,-0.424){2}{\rule{6.250pt}{0.102pt}}
\multiput(585.00,720.92)(17.028,-5.000){2}{\rule{3.125pt}{1.000pt}}
\multiput(615.00,715.71)(4.225,-0.424){2}{\rule{6.450pt}{0.102pt}}
\multiput(615.00,715.92)(17.613,-5.000){2}{\rule{3.225pt}{1.000pt}}
\multiput(646.00,710.69)(2.736,-0.462){4}{\rule{5.250pt}{0.111pt}}
\multiput(646.00,710.92)(19.103,-6.000){2}{\rule{2.625pt}{1.000pt}}
\multiput(676.00,704.69)(2.736,-0.462){4}{\rule{5.250pt}{0.111pt}}
\multiput(676.00,704.92)(19.103,-6.000){2}{\rule{2.625pt}{1.000pt}}
\multiput(706.00,698.69)(2.316,-0.475){6}{\rule{4.679pt}{0.114pt}}
\multiput(706.00,698.92)(21.289,-7.000){2}{\rule{2.339pt}{1.000pt}}
\multiput(737.00,691.68)(1.914,-0.481){8}{\rule{4.000pt}{0.116pt}}
\multiput(737.00,691.92)(21.698,-8.000){2}{\rule{2.000pt}{1.000pt}}
\multiput(767.00,683.69)(2.316,-0.475){6}{\rule{4.679pt}{0.114pt}}
\multiput(767.00,683.92)(21.289,-7.000){2}{\rule{2.339pt}{1.000pt}}
\multiput(798.00,676.68)(1.681,-0.485){10}{\rule{3.583pt}{0.117pt}}
\multiput(798.00,676.92)(22.563,-9.000){2}{\rule{1.792pt}{1.000pt}}
\multiput(828.00,667.68)(1.681,-0.485){10}{\rule{3.583pt}{0.117pt}}
\multiput(828.00,667.92)(22.563,-9.000){2}{\rule{1.792pt}{1.000pt}}
\multiput(858.00,658.68)(1.740,-0.485){10}{\rule{3.694pt}{0.117pt}}
\multiput(858.00,658.92)(23.332,-9.000){2}{\rule{1.847pt}{1.000pt}}
\multiput(889.00,649.68)(1.501,-0.487){12}{\rule{3.250pt}{0.117pt}}
\multiput(889.00,649.92)(23.254,-10.000){2}{\rule{1.625pt}{1.000pt}}
\multiput(919.00,639.68)(1.405,-0.489){14}{\rule{3.068pt}{0.118pt}}
\multiput(919.00,639.92)(24.632,-11.000){2}{\rule{1.534pt}{1.000pt}}
\multiput(950.00,628.68)(1.357,-0.489){14}{\rule{2.977pt}{0.118pt}}
\multiput(950.00,628.92)(23.821,-11.000){2}{\rule{1.489pt}{1.000pt}}
\multiput(980.00,617.68)(1.140,-0.492){18}{\rule{2.558pt}{0.118pt}}
\multiput(980.00,617.92)(24.691,-13.000){2}{\rule{1.279pt}{1.000pt}}
\multiput(1010.00,604.68)(1.180,-0.492){18}{\rule{2.635pt}{0.118pt}}
\multiput(1010.00,604.92)(25.532,-13.000){2}{\rule{1.317pt}{1.000pt}}
\multiput(1041.00,591.68)(1.056,-0.492){20}{\rule{2.393pt}{0.119pt}}
\multiput(1041.00,591.92)(25.034,-14.000){2}{\rule{1.196pt}{1.000pt}}
\multiput(1071.00,577.68)(1.018,-0.493){22}{\rule{2.317pt}{0.119pt}}
\multiput(1071.00,577.92)(26.192,-15.000){2}{\rule{1.158pt}{1.000pt}}
\multiput(1102.00,562.68)(0.921,-0.494){24}{\rule{2.125pt}{0.119pt}}
\multiput(1102.00,562.92)(25.589,-16.000){2}{\rule{1.063pt}{1.000pt}}
\multiput(1132.00,546.68)(0.865,-0.494){26}{\rule{2.015pt}{0.119pt}}
\multiput(1132.00,546.92)(25.818,-17.000){2}{\rule{1.007pt}{1.000pt}}
\multiput(1162.00,529.68)(0.800,-0.495){30}{\rule{1.882pt}{0.119pt}}
\multiput(1162.00,529.92)(27.095,-19.000){2}{\rule{0.941pt}{1.000pt}}
\multiput(1193.00,510.68)(0.698,-0.496){34}{\rule{1.679pt}{0.119pt}}
\multiput(1193.00,510.92)(26.516,-21.000){2}{\rule{0.839pt}{1.000pt}}
\put(250.0,750.0){\rule[-0.500pt]{7.468pt}{1.000pt}}
\put(1223.0,113.0){\rule[-0.500pt]{1.000pt}{91.301pt}}
\put(1223.0,113.0){\rule[-0.500pt]{51.312pt}{1.000pt}}
\end{picture}
\caption{The temperature dependence of the bag constant $B(t)$}
\end{figure}
\newpage
\begin{figure}
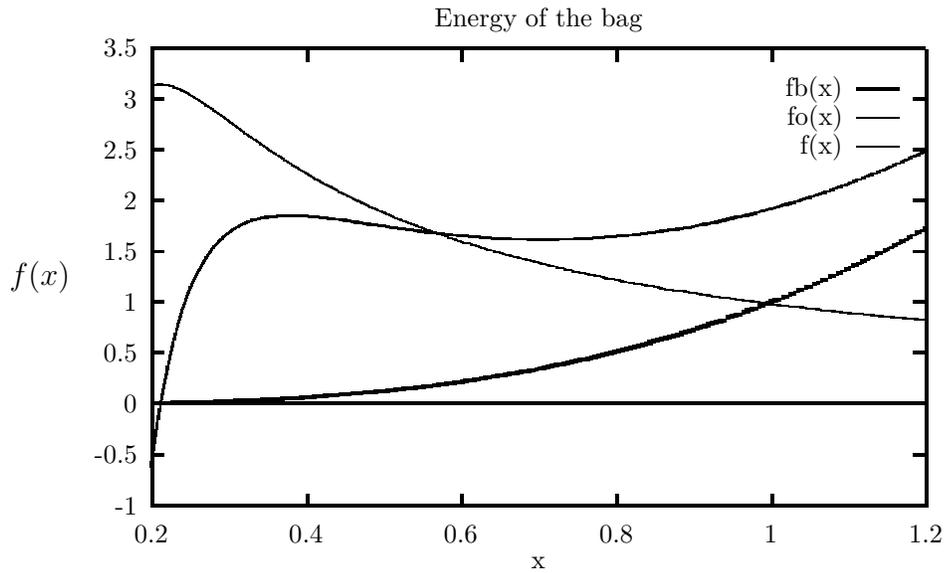

\input ball.tex
\caption{The effective energy of the bag loaded with quarks $f(x)$,
$f_b(x)$ - the energy of the empty bag, $f_o(x)$ - the energy of free quarks.}
\end{figure}

\end{document}